# Arc Discharge Carbon Nanoonions Purification by Liquid-Liquid Extraction

F.J. Chao-Mujica[1*], J.G. Darias-Gonzalez[1], L. Garcia-Hernandez[1], N. Torres-Figueredo[1,3], A. Paez-Rodriguez[2], L. Hernandez-Tabares[1], J.A.I. Díaz-Gongora[3], L.F. Desdin-Garcia[1*]

(1)-Centro de Aplicaciones Tecnológicas y Desarrollo Nuclear, Habana, Cuba, (2)- Centro de Estudios Avanzados de Cuba, Valle Grande, La Lisa, Habana, Cuba (3)-Centro de Investigación en Ciencia Aplicada y Tecnologia Avanzada. Instituto Politecnico Nacional, C. México, México.

*E-mail: chao@ceaden.edu.cu, desdin@ceaden.edu.cu

## Abstract

Carbon nanoonions are novel carbon nanoestructures that have potential applications in fields like electronics and chemical catalysis. Here we report a very simple but effective method of purifying carbon nanoonions produced by submerged arc discharge in water based on the water – toluene liquid-liquid extraction. Purified and non-purified samples were characterized by atomic force microscopy, high resolution transmission electronic microscopy and Brunauer-Emmett-Teller gas adsorption isotherms method. Microscopy results showed a good purification and allowed the assessment of the particles diameter distribution. Specific surface area was measured showing a great increment from $(14.7 \pm 0.3)$ m$^2 \cdot$g$^{-1}$ for the non-purified sample to $(170 \pm 3)$ m$^2 \cdot$g$^{-1}$ for the purified sample. Average particles diameter was also assessed from the adsorption isotherms; the diameter values obtained by the three techniques were in good agreement being between 20 to 30 nm.

## Introduction

Three decades has elapsed since the discovery of $C_{60}$[1]. However, carbon nanostructures remain a focal point of research. The versatility of carbon bonds allows the existence of multiple nanostructures[2]. Among the new members of the family of carbon allotropes are included the carbon nanoonions (CNO)[3] which are nanoparticles formed by concentric closed layers of sp$^2$ hybridized carbon atoms resembling the structure of an onion. The nomenclature of these structures in the literature is very variated, they are also named onion like carbon OLC, astralenes or multiwalled fullerenes. These structures present great variations in their size, morphology and layers structure though they can be easily classified into two great groups: polyhedral carbon nanoonions (pCNO), which have faceted, polyhedral shapes and the spherical carbon nanoonions (sCNO), which are almost spherical, well ordered and usually much smaller[4].

CNOs are promising nanomaterials with unique properties that make them ideal to a wide variety of applications[4] that's why many production methods have been widely studied[5]. Nevertheless, the large scale production of CNOs still faces up big technological and economic challenges. The methods which usually give the best results in terms of purity and consistency of the properties are not an economically viable alternative to produce large quantities of CNOs. The main expenses of these methods are associated with the requirement of vacuum systems, inert atmospheres, complex valve schemes and the use of catalysts which preparation is also a laborious task[6].

In addition to these methods, the submerged arc discharge (SAD) in water was proposed[7] and is being investigated[8] for CNOs fabrication. This method is a simpler and inexpensive alternative that bypasses the difficulties enumerated previously. However, this method produces a wide variety of carbon nanoparticles consisting of both sCNOs and pCNOs, wide-ranging multiwalled carbon nanotubes (MWCNT), graphenes and large amorphous and graphitic impurities. Therefore, this production method requires some sort of posterior purification/separation process to achieve the degree of uniformity specifically required for any given application. In literature two main approaches are followed; namely, the selective chemical destruction of unwanted structures and the physical separation of the particles according to their properties.

The selective chemical destruction of impurities is based on the premise that the amorphous carbon and other 'poorly ordered' or defective structures are more keen to react with oxidant agents than the wanted, 'well ordered' nanostructures. Chemical purification of CNOs have been reported by the annealing in air at temperatures over 400ºC[9] and using concentrated acids like nitric acid[9,10], nitric and sulphuric acid mixtures[11] and phosphotungstic acid[8]. Other oxidant agents like the hydrogen peroxide[12] have been also used in the purification of carbon nanotubes (CNT). These methods are of limited effectiveness, usually going in detriment of the yield of the desired particles and always introducing large amounts of oxygenated functionalities in their surfaces.

The physical separation, on the other hand, uses the differences in size, morphology and other physical properties of the particles in order to separate them. There's little literature on the use of this approach with CNOs but there are plenty of examples in the separation of other structures. Column chromatography it's being used for the separation of fullerenes since their discovery[13,14]. Ultracentrifugation have been used in the separation of CNTs[15–18]. Size exclusion chromatography have been reported in the separation of SWCNTs[19–22], MWCNTs[23] and more recently carbon quantum dots CQDs[24]. The selective dispersion in solvents or surfactants solutions of SWCNTs[18,25–27] and MWCNTs have also been reported[27].

Here we report a very simple physical method to effectively separate SAD produced CNOs from the other unwanted structures. This method is based on the water

– toluene liquid-liquid extraction and is thought to be readily scalable as part of any inexpensive SAD CNOs large production scheme.

Experimental

A submerged arc discharge (SAD) synthesis station developed by our team was used to synthesize the carbon nanoparticles. Our SAD system comprise three fundamental elements: an electrode gap micro positioning system controlled by the arc current measurement feedback, a ballast resistor for arc current stabilization and a data acquisition system to record the magnitude of the relevant physical parameters in a correlated way (see Figure 1). A detailed description of the experimental set up can be found in[28]. Spectroscopic pure graphite rods, with an apparent density of $\rho = 1.69 \pm 0.06$ g·cm$^{-3}$, were employed as electrodes (cathode: $\Phi = 12$ mm, length = 20 mm; anode: $\Phi = 6$ mm, length = 100 mm). The electrodes are arced under 8 cm of 1.2MΩ resistivity distilled water contained on a 5L double walled water-cooled stainless steel synthesis chamber. The CNO production was performed with a constant current of $I = 29.5 \pm 0.6$ A, a power of $P = 662 \pm 19$ W and a gap between the electrodes of 1 mm approximately.

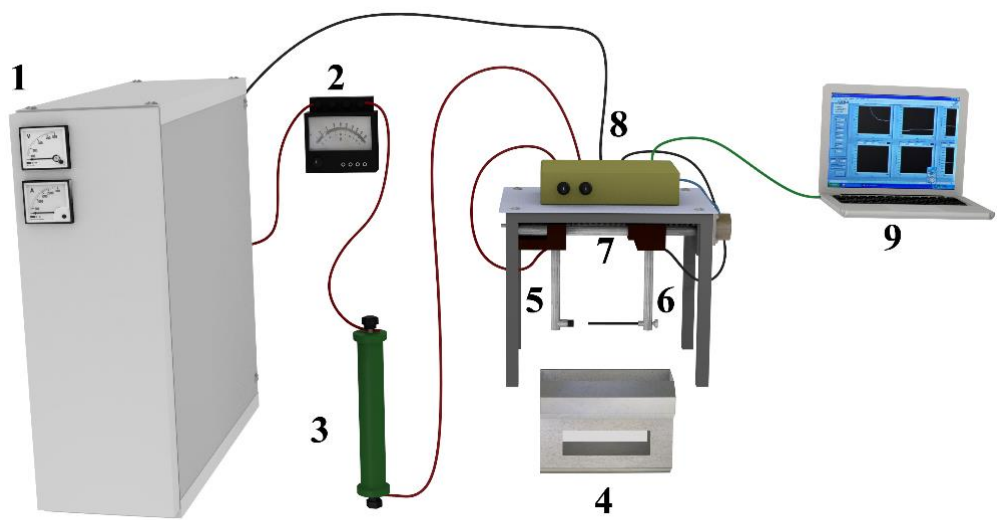

*Figure 1. SAD synthesis installation used to produce the carbon nanoparticles soot, power source (1), reference amperimeter (2), ballast resistor (3), cooled stainless steel synthesis chamber (4), cathode (5), anode (6), micro-positioning system (7), automated control unit (8) and PC for data acquisition (9).*

Synthesis products were allowed to rest for 24 hours. Then 50 ml of the water loaded with dispersed particles and floating material from the synthesis chamber were poured into a separation funnel. Then 50 ml of pure toluene, from Sigma-Aldrich, were also added to the funnel. The funnel was vigorously agitated for 5 minutes and then it was allowed to rest for other 10 minutes. After this process the funnel exhibits a darkened toluene phase, a much clearer aqueous phase and the apparition of interfacial black foam (see Figure 2). In this black foam is where the larger particles, like the macro and microscopic graphite particles and MWCNT remain. These particles get ¨stuck in the middle¨ being unable to disperse properly in the toluene phase due to their large size, weight and/or aspect ratio, nor to precipitate to the bottom of the funnel through the aqueous phase due to their high hydrophobicity.

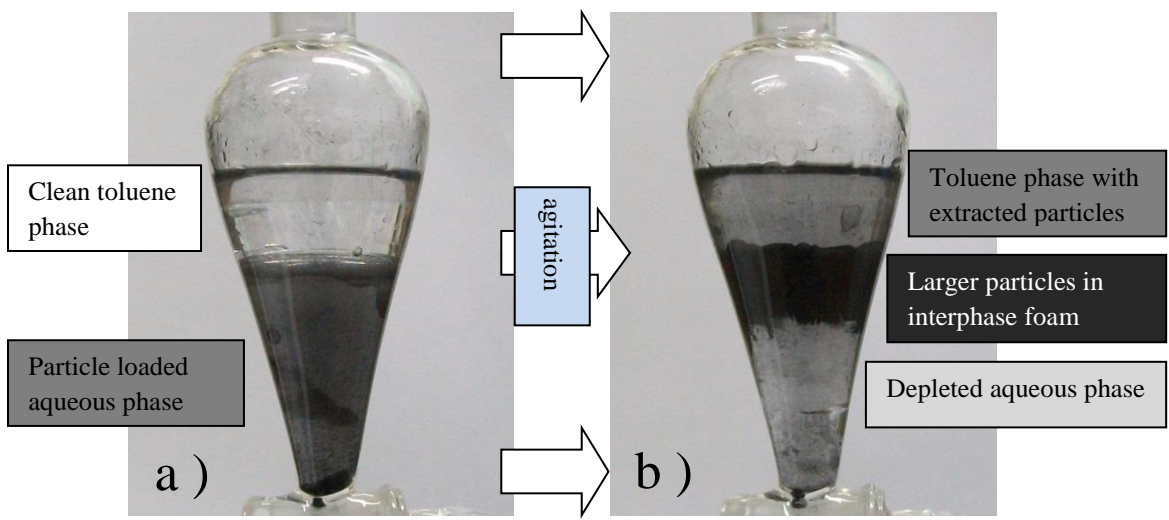

*Figure 2. Water-toluene extraction process, before (a) and after agitation (b).*

Afterwards the loaded toluene phase is retrieved from the funnel. Some of the as-obtained toluene loaded phase was separated to run the microscopy studies. The rest was slowly heated to completely evaporate the toluene and to obtain a dry black powder.

To make a comparison between the purified and the non-purified materials 50 ml of the as-synthesized water loaded with dispersed particles and floating material were also slowly and completely dried to powder.

The TEM images of the as-synthesized products were acquired with a JEOL JEM-2100 electron microscope and were processed with the software Gatan Microscopy Suite version 2.3[29]. The Atomic Force Microscopy (AFM) was carried out for the purified and the non-purified samples by the tapping method with a SPECTRA microscope from NT-MDT, using NSG-10 tips, of monocrystalline Si doped with Sb. The $N_2$ adsorption isotherms were measured, also for the purified and the non-purified samples, using ASAP 2050 equipment, Micromeritics technologies. Analysed samples were subjected to a prior activation process in temperature regime in vacuum at

120°C for 1 hour. The pressure values were recorded with an instrumental error of 1 µTorr and amount adsorbed an instrumental error of $0.0001 cm^3/g$.

## Results and Discussion

Both purified and non-purified samples were analysed using AFM (see

Figure 3). In the non-purified sample, it can be observed large clusters of graphite particles CNOs, MWCNT and other variated impurities. After the purification procedure just well dispersed CNOs are observed. The good dispersion in the purified sample allowed the analysis of individual particles giving that the CNOs average diameter D from the AFM data was $(30 \pm 9)$ nm.

Several HRTEM images were taken from the purified sample. From the TEM images the shell shaped concentric layered structure of the CNOs can be easily appreciated (see Figure 4). Processing of all of the TEM images taken gave an average CNOs diameter of $(20 \pm 8)$ nm. Despite that the results obtained by TEM and AFM are informative they are not representative since their images depict less than 1 pg of sample. Because of this, microscopy techniques alone are not capable of quantitatively evaluating the properties and purity of the typical inhomogeneous mixtures obtained by SAD.

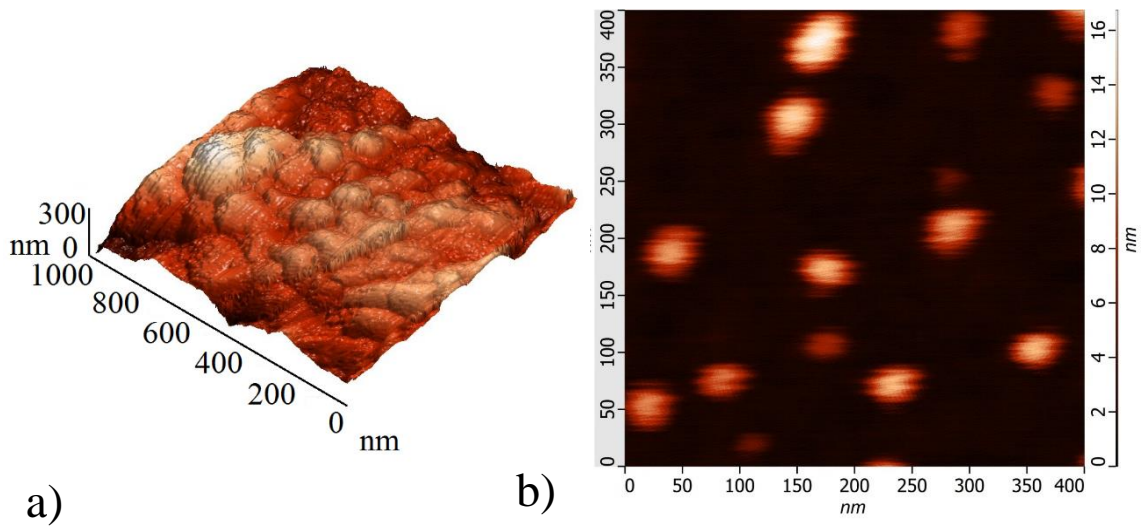

*Figure 3. AFM studies: a) Height diagram of the non-purified sample, b) Image of the purified sample showing well dispersed CNOs*

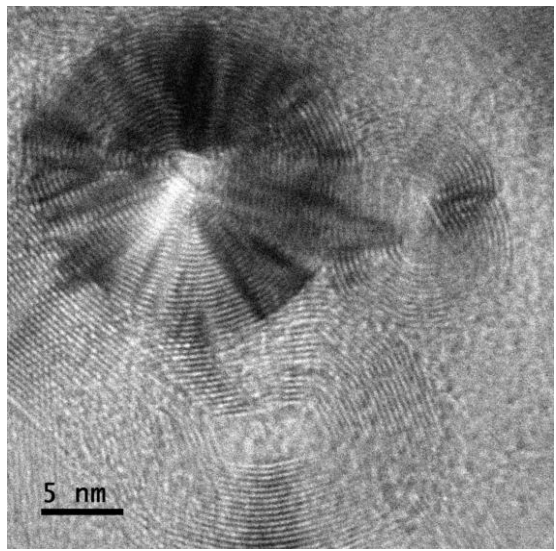

*Figure 4. Representative HRTEM image of a small CNOs cluster in the purified sample*

A bulk property like the specific surface area (SSA) is more suited to assess the effectiveness of the proposed purification process. The Brunauer-Emmett-Teller (BET) gas adsorption method with dinitrogen ($N_2$) as the adsorbing gas was used for the determination of the SSA of the samples. The SSA value for the as-produced materials was only (14.7 ± 0.3) m²·g⁻¹ while for the purified CNOs it reached (170 ± 3) m²·g⁻¹. If it is assumed that the sample consist of contacting nanoparticles close in size and with a smooth surface, then average diameter D of the particles can be estimated as[30]:

$$D = \frac{6}{\rho \cdot SSA} \quad (1)$$

Where $\rho$ is the density of the SSCNs. Using the value of $\rho$ (1.64 g·cm⁻³) for the CNOs reported elsewhere [13] it was calculated the average diameter D as 22 nm. The good agreement between the values of diameter obtained from TEM (20 ± 8) nm and AFM (30 ± 9) nm and the value estimated from the BET technique demonstrates that a high-quality purification was achieved.

Figure 5Also the adsorption and desorption isotherms were analysed (see Figure 5). The shape of the plot shows a typical type II behaviour[31]. This kind of isotherm is characteristic of non-porous or macroporous adsorbents with unrestricted monolayer-multilayer adsorption. This is in good agreement with the convex shape of CNOs and its non-porous surface. The observed hysteresis loop can be associated with type H3 loops. Type H3 loops are observed in aggregated particles forming slit like pores in the space between them, which again describes very well the purified CNO sample.

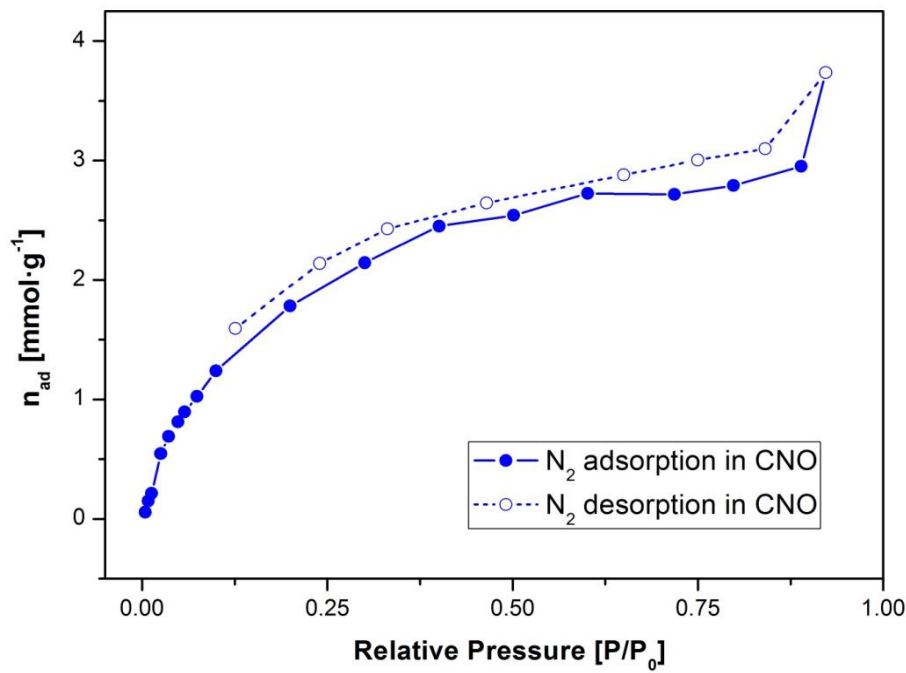

*Figure 5. The adsorption and desorption isotherms of $N_2$ on the purified CNOs sample.*

## Conclusions

A method based on liquid-liquid extraction for the purification of CNOs produced by SAD in water was presented. The main advantages of this method are its simplicity, both logistically and procedurally, and its easy scalability. A large increment in the SSA was achieved in the process from $(14.7 \pm 0.3)$ m$^2 \cdot$g$^{-1}$ to $(170 \pm 3)$ m$^2 \cdot$g$^{-1}$ for the purified sample. The average diameter of the CNOs assessed from the SSA 22 nm was similar to values measured by AFM $(30 \pm 9)$ nm and HRTEM $(20 \pm 8)$. The adsorption and desorption isotherms show that most of the usable surface area is readily accessible and that once adsorbed the gas can be quickly extracted. The relatively large SSA and high extraction rates of CNOs make them a promising material for many viable applications like the fabrication of electro-chemical double-layer capacitors EDLC and its use as chemical catalyst.

# Acknowledgements

The research was supported by AENTA-Cuba (PNUOLU-7-1, 2014) and by the Program for Basic Research (N@NO-C, 2015), MES-Cuba. L.F. Desdin-Garcia wishes to thank the SECITI (Edo. Mexico) and CLAF for generous support. N. Torres-Figueredo and J.A. Apolinar-Galicia acknowledge support from the CONACYT-Mexico. Authors thanks Prof. E. Reguera (CICATA-IPN, Legaria, Mexico, www.remilab.mx) for his support and encouragement.